# Large-area synthesis of ferromagnetic $Fe_{5-x}GeTe_2$/graphene van der Waals heterostructures with Curie temperature above room temperature

H. Lv,[1*] A. da Silva,[1] A. I. Figueroa,[2,8] C. Guillemard,[4] I. Fernández Aguirre,[2,3] L. Camosi,[2] L. Aballe,[4] M. Valvidares,[4] S. O. Valenzuela,[2,5] J. Schubert,[6] M. Schmidbauer,[7] J. Herfort,[1] M. Hanke,[1] A. Trampert,[1] R. Engel-Herbert,[1] M. Ramsteiner,[1*] and J.M.J. Lopes[1*]

1 - Paul-Drude-Institut für Festkörperelektronik, Leibniz-Institut im Forschungsverbund Berlin e.V, 10117 Berlin, Germany

2 - Catalan Institute of Nanoscience and Nanotechnology (ICN2), CSIC and BIST, Campus UAB, Bellaterra, 08193 Barcelona, Spain

3 - Universitat Autònoma de Barcelona, Bellaterra, Barcelona, 08193 Spain

4 - ALBA Synchrotron Light Source, Cerdanyola del Valles, 08290 Barcelona, Spain

5 - Institució Catalana de Recerca i Estudis Avançats (ICREA), 08010 Barcelona, Spain

6 - Peter Grünberg Institut (PGI-9), Forschungszentrum Jülich, 52425 Jülich, Germany, and JARA-Fundamentals of Future Information Technology, Jülich-Aachen Research Alliance

7 - Leibniz-Institut für Kristallzüchtung (IKZ), 12489 Berlin, Germany

8 – Current Address: Departament de Física de la Matèria Condensada, Universitat de Barcelona, and Institut de Nanociència i Nanotecnologia, Universitat de Barcelona, 08028 Barcelona, Spain

**ABSTRACT**

Van der Waals (vdW) heterostructures combining layered ferromagnets and other two-dimensional (2D) crystals are promising building blocks for the realization of ultra-compact devices with integrated magnetic, electronic and optical functionalities. Their implementation in various technologies depends strongly on the development of a bottom-up scalable synthesis approach allowing to realize highly uniform heterostructures with well-defined interfaces between different 2D layered materials. It also requires that each material component of the heterostructure remains functional, which ideally includes ferromagnetic order above room temperature for 2D ferromagnets. Here, we demonstrate large-area growth of $Fe_{5-x}GeTe_2$/graphene heterostructures achieved by vdW epitaxy of $Fe_{5-x}GeTe_2$ on epitaxial graphene. Structural characterization confirmed the realization of a continuous vdW heterostructure film with a sharp interface between $Fe_{5-x}GeTe_2$ and graphene. Magnetic and transport studies revealed that the ferromagnetic order persists well above 300 K with a perpendicular magnetic anisotropy. In addition, epitaxial graphene on SiC(0001) continues to exhibit a high electronic quality. These results represent an important advance beyond non-scalable flake exfoliation and stacking methods, thus marking a crucial step toward the implementation of ferromagnetic 2D materials in practical applications.

---

[*] Corresponding authors: lv@pdi-berlin.de, ramsteiner@pdi-berlin.de, lopes@pdi-berlin.de

**INTRODUCTION**

Two-dimensional (2D) ferromagnetic materials are anticipated to play a crucial role in the development of future spin-based devices (e.g., spin-valves, random-access magnetic memories) with high energy efficiency and ultra-compact dimensions.[1–3] Moreover, combining them with other layered materials such as graphene and 2D semiconductors (e.g., $WS_2$ and $MoSe_2$) into van der Waals (vdW) heterostructures is a promising pathway for the realization of novel devices with integrated magnetic, optical, and electronic functionalities.[4] In this case, besides the individual properties of each 2D crystal, proximity-induced coupling effects across the vdW gap offer unique opportunities to tailor properties via a layering design.[4,5]

In order to enable a broad range of applications, it is critical to identify 2D ferromagnets exhibiting a Curie temperature ($T_C$) at least at room temperature.[2,6] Among a vast selection of 2D materials with magnetic order investigated recently,[6] the $Fe_{5-x}GeTe_2$ (FGT) metallic system (with x usually between 0 and 0.4)[7–10] has been found to offer a great prospect, as it has been reported to exhibit a $T_C$ around 300 K.[7,10,11] FGT is an itinerant ferromagnetic metal[11] with magnetic properties that can be tuned via electrostatic gating,[12] strain or doping when interfacing a topological insulator,[13] as well as elemental doping.[14,15] Similar to other 2D Fe-Ge-Te compounds like $Fe_3GeTe_2$[16] and $Fe_4GeTe_2$[17] with below room temperature $T_C$, it possesses a layered structure with each single layer being formed of Fe and Ge 2D slabs encapsulated by layers of Te.[7,8] The weak vdW interaction between interlayers permits to exfoliate micrometer-sized flakes from FGT bulk single crystals and to fabricate vdW heterostructures combining FGT and other 2D crystals via sequential flake stacking.[18–20] Remarkably, such approach has been used to realize FGT/graphene spin-valve devices demonstrating room temperature operation.[18] Nevertheless, despite its great utility for fundamental studies and testing device concepts, exfoliation-based methods are not scalable and difficult to integrate into established device fabrication schemes.

Achieving bottom-up synthesis of vdW heterostructures comprising FGT and other 2D materials would overcome a roadblock towards implementing these materials into future devices. In this work, we demonstrate the synthesis of large-area FGT/graphene vdW heterostructures exhibiting ferromagnetic order persisting well above 300 K and perpendicular magnetic anisotropy. For this, vdW epitaxy of FGT films on single-crystalline graphene/SiC(0001) templates was realized by molecular beam epitaxy (MBE), a method which has recently been employed for the growth of FGT on substrates such as $Al_2O_3(0001)$,

WS$_2$(001)/Al$_2$O$_3$(0001), and SrTiO$_3$(111).[9,13,21] Importantly, the underlying graphene layers exhibit a sharp interface to the FGT and conserve their high electronic quality, which are key ingredients for the future development of devices in which interfacial effects can be used to control properties and functionalities.

## RESULTS AND DISCUSSION

### Structural properties

MBE was employed to grow ~15 nm thick FGT films on epitaxial graphene on SiC(0001) templates synthesized by surface graphitization.[22] Rutherford backscattering spectrometry (RBS) analyses (see Supporting Information, Figure S1) revealed the films to exhibit an average composition close to x = 0, namely Fe$_{4.8}$GeTe$_2$. This is similar to what has been reported for bulk single crystals synthesized by chemical vapor transport.[7,23] *In-situ* growth monitoring by reflection-high-energy-electron-diffraction (RHEED) showed a pattern composed of isolated streaks which indicates growth of a film with a smooth surface (Figure 1a, bottom panel). The position of the streaks with respect to the pattern measured for epitaxial graphene on SiC(0001) before growth [for SiC(0001), the in-plane lattice parameter is 3.08 Å][24] attested the evolution to a larger in-plane lattice constant around 4.06 Å, as expected for FGT formation.[9]

To study the in-plane structure in more detail, synchrotron-based grazing incidence diffraction (GID) was employed. Figure 1b shows the in-plane intensity distribution plotted in hexagonal coordinates according to the symmetry of the SiC(0001) substrate. This map was obtained for the 15 nm FGT film covered by a 5 nm thick Te capping layer, which was used to reduce FGT surface oxidation upon prolonged air exposure. For convenience, we use the four-component vector notation for hexagonal symmetry: (*H K. L*) = (*H K* –[*H+K*] *L*). The only substrate contribution within the investigated area in reciprocal space appears as a single ($2\bar{1}.0$)-type reflection. The other most intense and prominent diffraction features originate from the FGT film. There are two different sets of reflections, namely FGT(11.0) and (22.0), and FGT(H0.0) with [*H*=1..3]. They both refer to orientations with their c-axes along the surface normal. The three most intense diffraction features at (11.0), (22.0) and (30.0) reveal a FGT in-plane lattice parameter of 4.049(13) Å. This is in general agreement with the value obtained by RHEED, as well as literature values for Fe$_{5-x}$GeTe$_2$ (0 < x ≤ 0.4) bulk crystals[7,8,10] and thin films.[9,13]

The intensity modulation along the FGT(11.0) ring exhibits a maximum in the azimuth which is aligned along the SiC[$2\bar{1}$.0] in-plane substrate direction. This indicates a preferential in-plane orientation, i.e. the (11.0) net planes of FGT and SiC are parallel, as we previously observed for Fe$_3$GeTe$_2$ films grown on graphene.[25] Nevertheless, the FGT arcs maxima are broader than those of Fe$_3$GeTe$_2$. For the arc crossing the (11.0) reflection, a full width at half maximum of 8.9° (inset, Figure 1b) is found, whereas for Fe$_3$GeTe$_2$ this value is about 4.7°.[25] Even though this relatively large value is not uncommon for epitaxially grown vdW materials,[26–28] it reveals that the in-plane mosaicity in the FGT film is clearly larger than that measured for the Fe$_3$GeTe$_2$ one (grown with virtually identical conditions, except the Fe flux). This difference might be associated with the higher energies required for the stabilization of layered Fe-Ge-Te phases exhibiting an Fe composition higher than 3.[17] The existence of in-plane mosaicity can also be inferred from the rather diffuse streaks of the RHEED pattern (Figure 1a, bottom panel). Similar RHEED patterns were reported by Ribeiro *et al.*[9] for $Fe_5GeTe_2$ films grown by MBE on $Al_2O_3$(0001) substrates. Interestingly, they could observe an evolution from diffuse to sharp RHEED streaks by performing post-growth annealing of the films at a temperature of 550 °C under Te flux. This suggests that post-growth thermal treatment as well as higher growth temperatures might improve in-plane order for FGT films grown on graphene. This will be the topic of further investigations. There are two more ring-like contributions detectable in the map and in the attached line profile taken at $K$=0, which cannot be related to FGT. These are most probably due to a smaller fraction of tetragonal FeTe agreeing very well with FeTe(200) and (220) in-plane lattice spacing. Interestingly, the c-axis of tetragonal FeTe points along the direction perpendicular to the surface as well. Finally, a sequence of low-index poly-rings with different lattice spacings is also detected, which is originating from the topmost Te capping layer.

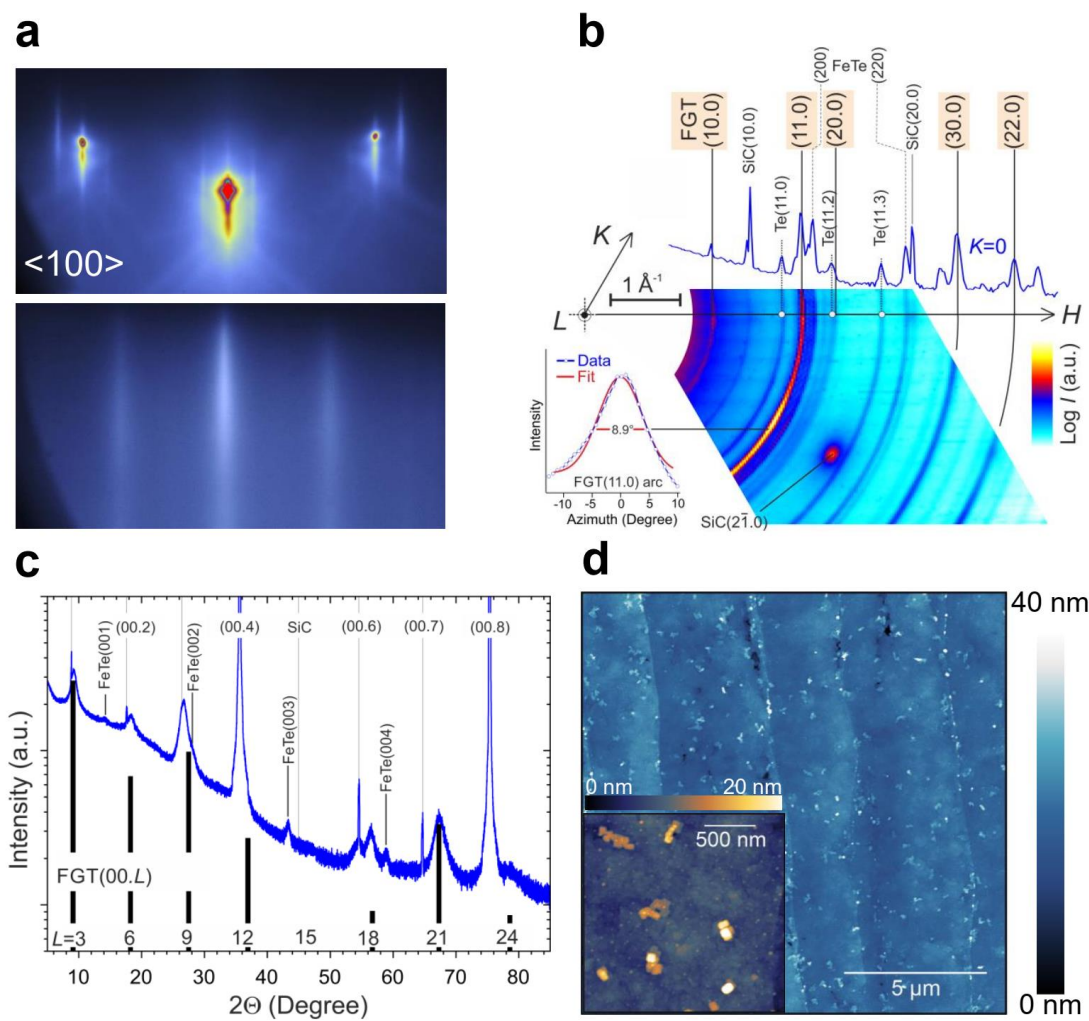

**Figure 1.** Structural characterization of 15 nm thick FGT films on graphene. a) *In-situ* RHEED patterns taken perpendicular to the SiC<100> direction, before (upper panel) and after FGT growth (lower panel). b) GID in-plane reciprocal space map. It contains the SiC (2$\bar{1}$.0) reflection, as well as contributions along particular rings (corresponding to discrete in-plane lattice parameters) which can be well attributed to FGT, FeTe, and the Te capping layer. The inset image (left side, bottom) is an azimuthal intensity profile intersecting the maximum of the (11.0) FGT arc. c) Out-of-plane omega/2theta scan for an uncapped film containing a series of FGT peaks, whose intensities and positions match well with calculated ones. It also shows contributions from FeTe and the SiC substrate. d) AFM height image taken for an uncapped film. The inset depicts a higher magnification image where it is possible to better visualize the nanostructures present at the surface.

Complementary to GID we have performed a laboratory-based omega/2theta measurement at an x-ray wavelength of 1.54056 Å, which probes exclusively out-of-plane features of the layer, see Figure 1c. A set of FeTe reflections confirms the presence of tetragonal FeTe with its out-of-plane orientation along SiC[00.1] as it could already be anticipated from the GID in-plane data. A further set of eight subsequent FGT(00.$L$) reflections with $L$ = 3, 6, .., 24 proves (in conjunction with the presented in-plane data) the formation of FGT with a $Fe_5GeTe_2$ structure,[8] since they fit well to the position and intensity ratios of the numerically derived powder pattern (black bars). Based on the positions of the contributions at $L$=3, 6, 18 and 21 we have deduced an out-of-plane lattice parameter of 29.16(9) Å, in agreement with previous reports for bulk crystals and thin films.[7–10,13] The formation of [00.1]-oriented $Fe_3GeTe_2$ can be excluded as it would result into a clearly different diffraction pattern.[16,29] Note that these measurements were performed for uncapped FGT fims, which explains the absence of contributions related to pure Te.

The surface morphology was probed by atomic force microscopy (AFM). Figure 1d shows a typical AFM height image for an uncapped FGT film. Its coverage appears mainly continuous over the micrometer-sized graphene/SiC terraces, with exception of some holes appearing close to surface steps. These substrate regions are known to exhibit few-layer thick graphene ribbons and patches, whereas the terraces are mostly covered in monolayer- and, to a lesser extent, bilayer-thick graphene.[28,30] A similar effect was observed in $Fe_3GeTe_2$ grown on graphene/SiC(0001),[25] which we associated to the lower surface reactivity of the thicker graphene areas relative to that of the terraces,[22] leading to a lower FGT growth rate near the steps. Moreover, it is possible to observe nanoislands with tetragonal shapes scattered over the surface (inset, Figure 1d). Based on the GID and XRD findings, we suggest that they are composed of tetragonal FeTe,[31] forming due to phase segregation during growth or during subsequent sample cooling. In spite of these unwanted surface effects – which are expected to be mitigated by further refinement of both the FGT and graphene growth protocols – it is reasonable to conclude that the FGT film offer a good surface homogeneity over large areas. The root-mean-square roughness (RMS) for different surface regions (free of FeTe nanoislands) varies between 0.8 and 1.1 nm, which is close to the thickness of 1 monolayer of $Fe_5GeTe_2$ [a single layer corresponds to 1/3 of the unit cell with $c$ = 29.16(9) Å]. It is important to note that the AFM investigations discussed here were conducted on non-capped FGT films. Although oxidation in FGT[32] is not as strong as in 2D magnetic halides,[33] surface oxidation taking place within a few hours after air exposure might increase the surface roughness.

The structure of the FGT/graphene heterostructure was also investigated by scanning transmission electron microscopy (STEM). Figure 2a depicts a high-angle annular dark-field (HAADF) STEM image of the FGT film grown on graphene/SiC(0001). It is possible to observe the layered structure of the film, in that single layers are clearly separated by vdW gaps, in agreement with previous TEM characterization of FGT bulk crystals and thin films.[7,9] Most of the film is formed of well-oriented layers with the expected FGT single layer thickness (~1 nm, see Figure 2b). This value agrees with the $c$ lattice constant obtained from the XRD data. Stacking and rotational faults are also evident. Strikingly, STEM images reveal that, although the overall vdW-layered structure is preserved, single layers with a thickness larger than 1 nm are also present. Figure 2c, which is an enlarged area from an interface region in Figure 2a, reveals some layers as thick as 2 nm. To the best of our knowledge, the experimental realization of Fe-Ge-Te layered phases in which the thickness of a single layer is larger than that of $Fe_5GeTe_2$ has not been reported. In a simple picture, such a thickness increment is anticipated if one considers that "filling" the 2D backbone structure of $Fe_5GeTe_2$ with more Fe atoms will result in the formation of additional slabs of Fe, following the same trend as experimentally observed when transitioning from $Fe_3GeTe_2$ to $Fe_5GeTe_2$ (for which the single-layer thickness evolves from ~0.8 nm to ~1.0 nm).[7,8,16,34] In fact, Liu *et al.*[35] used first-principle calculations to study different Fe-Ge-Te layered phases including $Fe_6GeTe_2$ and $Fe_7GeTe_2$. They predicted them to exhibit a thickness for a single layer of around 1.1 and 1.2 nm, respectively. Our XRD data (Figure 1c) do not directly confirm the formation of such phases, which shows that they are not dominant in the film. However, the shift of the (00.9) reflection towards a lower angle (in comparison to the value calculated for $Fe_5GeTe_2$), as well as the existence of an additional, unidentified reflection closed to (00.18) [superimposed to SiC (00.6)], could be related to the existence of layered phases with larger $c$ lattice constants.

Even though at the present stage we cannot unambiguously identify their composition and structure, the existence of single layers with thickness larger than that of $Fe_5GeTe_2$ is a strong evidence for the formation of new Fe-Ge-Te layered phases with Fe composition exceeding 5, which are possibly metastable.[17] This illustrates the great potential of vdW epitaxy via MBE for the large-scale realization of metastable phases of 2D magnets, similar to what has been observed for 2D semiconducting dichalcogenides.[36,37] Further investigations are needed in order to identify the reasons for the formation of such phases and to controllably realize them. Finally, note that the average RBS composition obtained for our films ($Fe_{4.8}GeTe_2$) does not pose a contradiction to the existence of layers with higher Fe concentration. It is known that $Fe_{5-x}GeTe_2$ single layers are usually Fe-deficient with $0 < x \leq 0.4$.[8,10,38,39] (x can be even close to 1 if the $Fe_4GeTe_2$ structure stabilized by Seo *et al.*[17] is considered, which interestingly also exhibits a single layers thickness around 1 nm). Fe deficiency taking place in the "normal" FGT layers could in principle compensate for the higher amount of Fe present at the less abundant, thicker layers.

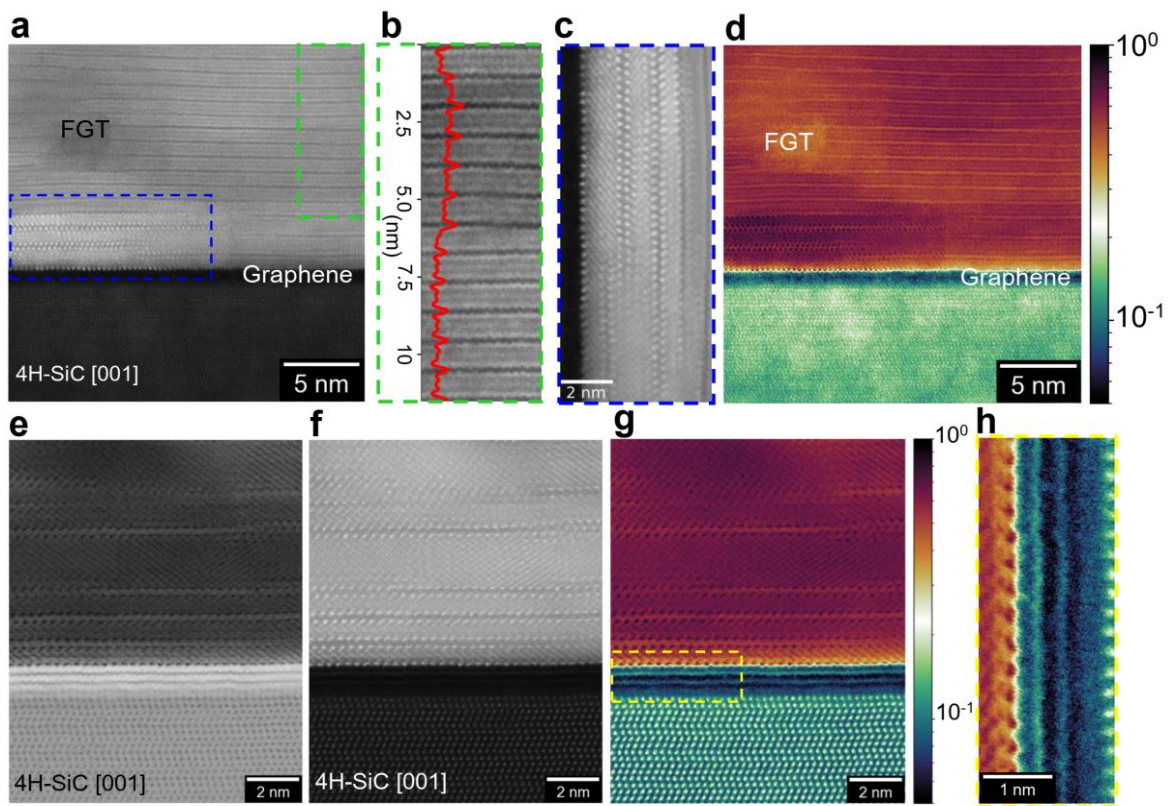

**Figure 2.** STEM of the 15 nm thick FGT film on graphene/SiC(0001). a) HAADF-STEM image of a region showing the FGT film on graphene. b) Magnified region from green dashed rectangle in a) showing the homogeneous layer formation and its line profile (in red). The interlayer distance is 0.97 (±0.10) nm. c) Another enlarged area (blue dashed rectangle) from a), showing the formation of single layers with different thicknesses. d) The same image as a) rendered using a custom colormap to aid visualization. e) Bright field-STEM image showing another region of the sample where the graphene layers are better resolved. f) HAADF-STEM image of the same area in e). g) colormap version of the same area in e) to aid visualization of the graphene layers. h) Enlarged region of the yellow rectangular box in g).

Another very important feature observed in the STEM images is the sharp nature of the FGT/graphene interface. Figure 2d depicts the same area as in Figure 2a, rendered using a custom colormap to aid the visualization of the graphene layers, since the HAADF-STEM images provide an atomic number (Z) contrast. The continuous graphene layers are observed (dark green color) directly interfacing the crystalline FGT film (in orange). A more detailed view of the FGT/graphene interface, obtained for another sample region, is provided in Figures 2e-h. In agreement with previous reports,[22] they confirm the existence of a bilayer thick graphene (with an interlayer distance of 0.34 nm), and the carbon-rich buffer layer very close to SiC.

**Magnetic and electric properties**

The magnetic and electric properties were investigated via magneto-transport measurements, superconducting quantum interference device (SQUID) magnetometry, X-Ray absorption spectroscopy (XAS), and X-ray magnetic circular dichroism (XMCD). These measurements were performed on pieces cut from the same 1 cm$^2$ sample consisting of a ~ 15 nm FGT film (capped with 5 nm Te) grown on graphene/SiC(0001).

Figure 3a displays the Hall resistivity $\rho_{xy}$ measured via magneto-transport during upward and downward sweeps of an external magnetic field $H$, where the arrows indicate the field sweeping direction. The linear background is caused by the ordinary Hall effect (OHE) with the negative slope observed for the whole temperature range suggesting electron-like transport behavior. However, the measured OHE results from parallel conduction through the FGT and graphene transport channels exhibiting hole- and electron-like transport characteristics, respectively.[25] Consequently, the sign as well as the absolute value of the OHE slope depend on the relative conductivities of the FGT and graphene films. The extracted effective carrier density of 1.0 × 10$^{14}$ cm$^{-2}$ and the sheet resistance of 176 Ω/sq (at room temperature) in our FGT/graphene heterostructure are close to the values obtained for the previously studied Fe$_3$GeTe$_2$/graphene heterostructure (7 × 10$^{13}$ cm$^{-2}$ and 150 Ω/sq, respectively).[25] The observed hysteresis loop superimposed onto the OHE at each temperature originates from the anomalous Hall effect (AHE), its signal strength being proportional to the out-of-plane magnetization. The nearly square-shaped loops and the finite coercivity observed for temperatures up to 360 K provide clear evidence for ferromagnetism well above room temperature with a robust perpendicular magnetic anisotropy (PMA). For the determination of the Curie temperature, we adopt the commonly used method based on the Arrott plot:[17,40] $\rho_A^2$ is plotted against $\mu_0 H/\rho_A$, where $\rho_A$ is the AHE part of the Hall resistivity after subtracting the linear background of the OHE contribution (see also Supporting Information, Figure S2). From the linear fit in the high-field ranges of the Arrott plots in Figure 3b, the intercepts at the $\rho_A^2$ axis are extracted for each temperature. As shown in Figure 3c, these intercepts exhibit a linear temperature dependence in accordance with Arrott's theory.[40] Finally, a Curie temperature of $T_C \approx 390$ K is extracted from the extrapolation to zero intercept by a linear fit (red line in Figure 3c). The same $T_C$ is obtained from the temperature decays of the remanent AHE resistivity and the coercive field (Supporting Information, Figure S3). This value is significantly higher than those previously reported for single crystals (or flakes) as well as large-area films of $Fe_3GeTe_2$ ($T_C \approx 220$ K),[25,29,41,42] $Fe_4GeTe_2$ ($T_C \approx 270$ K),[17] and $Fe_{5-x}GeTe_2$ ($0 < x \leq 0.4$, $T_C \approx 280 - 320$ K). [7,9,21,39,43] Only the transition temperature recently reported for MBE-grown $Fe_{5-x}GeTe_2/Bi_2Te_3$ heterostructures ($T_C$ as high as 570 K) surpasses this value.[13]

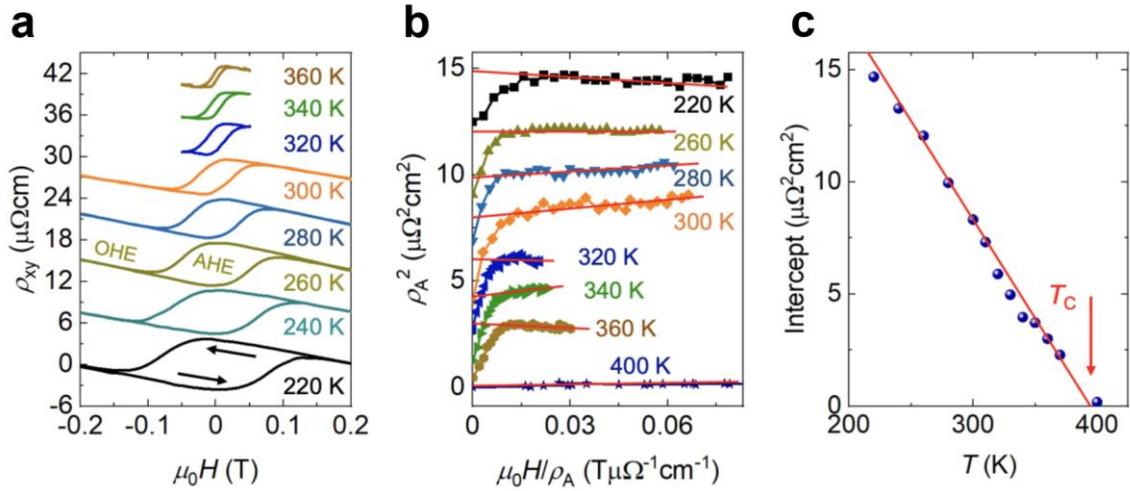

**Figure 3**. Curie temperature determination using magnetotransport measurements for a Te capped, 15 nm thick FGT film on graphene/SiC(0001). a) Hall resistivity $\rho_{xy}$ measured during upward and downward sweeps of an external magnetic field at various temperatures with the sweeping directions indicated by arrows. The square-shaped hysteresis loops result from the AHE. b) Arrott plot ($\rho_A^2$ against $\mu_0 H/\rho_A$) extracted from AHE data, where $\rho_A$ is the AHE part of the Hall resistivity. The linear dependencies in the large field range are indicated as red lines. c) Intercepts at the $\rho_A^2$ axis of Arrott plots as a function of temperature. The Curie temperature $T_C$ is determined from the extrapolation to zero intercept from a linear fit (red line) of the data.

The high $T_C$ value approaching 400 K is in agreement with a recent theoretical prediction for the $Fe_5GeTe_2$ phase,[17] however it differs from other experimental reports for $Fe_{5-x}GeTe_2$. This could be related to the existence of thicker single vdW FGT layers, as observed by STEM. It is anticipated that a gradual increase of the Fe content within individual layers will result in a larger number of nearest Fe neighbors per Fe atom, enhancing the spin-pair exchange interaction and consequently the Curie temperature. In fact, recent theoretical investigations predicted PMA and $T_C$ values of 450 K and 570 K for layered $Fe_6GeTe_2$ and $Fe_7GeTe_2$, respectively.[35] Therefore, in the case of our sample, the existence of similar phases with Fe composition higher than 5 could in principle lead to an effective $T_C$ value that is superior to that of pure $Fe_{5-x}GeTe_2$. Nevertheless, in order to verify this hypothesis, future growth experiments will be conducted aiming at the stabilization, and isolation, of such phases. Finally, note that the tetragonal FeTe phase detected by GID and XRD (Figure 1b,c) and correlated to the surface nanoislands observed by AFM (Figure 1d), is not expected to contribute to the observed ferromagnetism due to its antiferromagnetic nature (Néel temperature around 70 K).[31]

The magnetic properties of the FGT film were further investigated by SQUID magnetometry. Figure 4 displays the remanent magnetization obtained for both out-of-plane (OP) and in-plane (IP) configurations as a function of temperature (a background signal caused by the SiC substrate has been subtracted; Supporting Information, Figure S4). The clearly larger remanent magnetization in the OP configuration confirms the PMA in our FGT films. Furthermore, it is clear from the decay of the OP remanent magnetization that ferromagnetic order persists well above room temperature, in agreement with the result extracted from the AHE measurements. The Fe atoms on the five inequivalent lattice sites in $Fe_5GeTe_2$ are expected to exhibit rather different magnetic moments.[7,38] The average magnetic moment per Fe atom obtained from the SQUID data measured at 2K is 1.37 $\mu_B$, considering the average Fe concentration of 4.8 as obtained by RBS measurements. This result falls into the range between 0.8 and 2.6 $\mu_B$ per Fe atom covered by reported experimental values.[7]

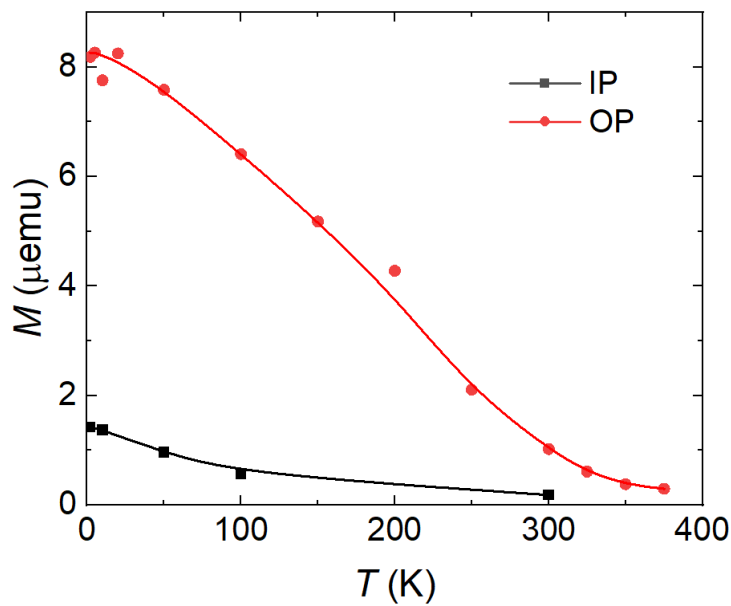

**Figure 4.** Remanent magnetization $M$ obtained by SQUID magnetometry as a function of temperature for in-plane (IP) and out-of-plane (OP) configurations on the same sample piece of a Te-capped, 15 nm thick FGT film on graphene/SiC(0001) with a FGT volume of about $1.07 \times 10^{-4}$ mm$^3$. The background signal originating from the SiC substrate was subtracted (Figure S4, Supporting Information).

XAS and XMCD measurements were employed to investigate the atomic scale magnetic properties.[44] Figure 5 shows the remanence of the dichroic signal at $B$ = 0 T (normalized by its value at B = 6 T), for both grazing (20º from the sample plane) and normal incidence geometries, as a function of temperature. At low temperature, the sample has a PMA, as evidenced by the larger remanence in the perpendicular direction (red triangles). However, above T = 220 K, the measured magnetization undergoes an in-plane reorientation transition. The spontaneous magnetization persists above room temperature (Supporting Information, Figure S6 and S7). We performed a sum-rules analysis on the XAS-XMCD spectra at T = 3 K to extract the atomic magnetic moments of Fe atoms (Supporting Information, Figure S5). We obtained an effective total magnetic moment in normal incidence (i.e. not corrected from the intra-atomic dipole operator $T_z$) $M_{eff} = 1.13 \pm 0.15\ \mu_B$/Fe atom, in reasonable agreement with the value obtained by SQUID. The lower value found from this analysis might be due to the fact that not all Fe atoms probed in XAS contribute to the magnetic signal obtained by XMCD (e.g., those part of antiferromagnetic FeTe nanoislands do not contribute), thus this value is an underestimation of the actual Fe magnetic moment in FGT. This analysis also allowed us to extract the orbital moments measured in normal incidence $m_L^\perp = 0.03\ \mu_B$ /Fe and the difference between perpendicular and in plane orbital moments $\Delta m_L = m_L^\perp - m_L^\parallel$ that is proportional to the macroscopic anisotropy in itinerant magnetism materials according to

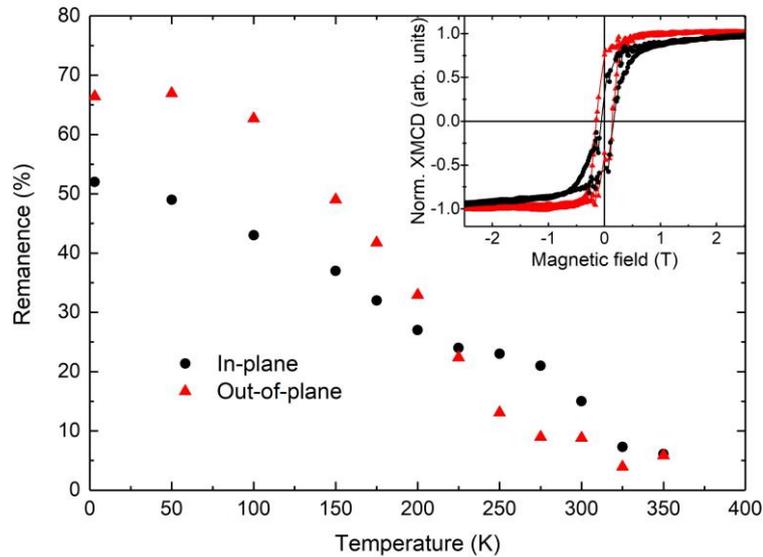

**Figure 5.** Temperature dependence of the Fe $L_3$ XMCD maximum in remanence (0 T), normalized by its value in saturation (6 T). The external magnetic field was applied along the direction perpendicular to the sample surface (out-of-plane, red triangles) and at 20º from the sample plane (in-plane, black circles). Inset: Hysteresis loops recorded at the Fe $L_3$ edge at 3 K for both configurations, where data have been normalized by the value in saturation (6 T).

Bruno's model.[45] We found $\Delta m_L < 0.01\ \mu_B$, consistent with the small anisotropy observed in normal (red curve) and grazing (black curve) hysteresis loops shown in the inset of Figure 5.

A $T_C$ above room temperature, as observed in XMCD, is in good agreement with literature values for $Fe_{5-x}GeTe_2$ with x ranging from 0 to 0.3.[9,10,21,39] Certain discrepancy observed between XMCD, SQUID, and magneto-transport results, in terms of remanence, anisotropy and $T_C$ may be attributed to the fact that the different experiments probe different parts of the FGT film. While SQUID probes the whole film, the detection in XMCD is restricted to the topmost FGT layers (~ 5 nm). As observed by STEM (Figure 2a), this area is formed exclusively of single layers with the interlayer spacing expected for $Fe_{5-x}GeTe_2$ with x close to 0. Also, the grazing geometry measured in XMCD does not correspond to a purely in-plane component of the magnetization, but instead it has a contribution from the out-of-plane one. For the magneto-transport (AHE) measurements, the actual current distribution between the different FGT components as well as the graphene film is essential but difficult to determine.

Finally, it is important to verify the functionality of the graphene film after FGT synthesis. Figure 6 displays the longitudinal magnetoresistance ($R_{xx}$) of the FGT/graphene stack measured at 4.3 K under a perpendicular magnetic field. The apparent hysteresis is caused by the magnetization switching in the FGT film, while the $H^2$-dependent positive magnetoresistance (red line) and the peak induced by weak localization at zero field (pink square area) are typical characteristics of the carrier transport in graphene.[46–50] This observation reveals the occurrence of strong quantum interference effects for the electron motion, demonstrating that the high quality of the graphene film is maintained during the overgrowth by FGT. In the presence of a magnetic field, the strength of weak localization in the 2D case can be described by the Hikami-Larkin-Nagaoka (HLN) equation:[51]

$$\Delta\sigma_{xx} = \alpha \frac{e^2}{\pi h}\left[\psi\left(\frac{1}{2} + \frac{H_\phi}{H}\right) - ln\left(\frac{H_\phi}{H}\right)\right]$$

where $\Delta\sigma_{xx}(H) = \sigma_{xx}(H) - \sigma_{xx}(0)$ is the corresponding change in the longitudinal conductivity. Here, ψ is the Digamma function, $H_\phi$ the phase coherence characteristic field, $h$ the Planck constant and $e$ the elementary electron charge. In the low field range (0.2 to 0 T during down-sweep of the external field to avoid the influence of magnetization switch), our data can be well fitted by the HLN equation using $\alpha$ and $H_\phi$ as free parameters as shown by the pink line in Figure 6b. The obtained $\alpha$ = 0.45 is in accordance with previously reported values and, therefore, confirms the quantum origin of the resistance up-turn.[51] Based on the extracted $\mu_0 H_\phi$ = 3.67 mT, a phase coherence length of $l_\phi$ = 211 nm is deduced from $\mu_0 H_\phi = h/8\pi e l_\phi^2$. The obtained $l_\phi$ is comparable to a previously reported value for epitaxial graphene on SiC.[49,52]

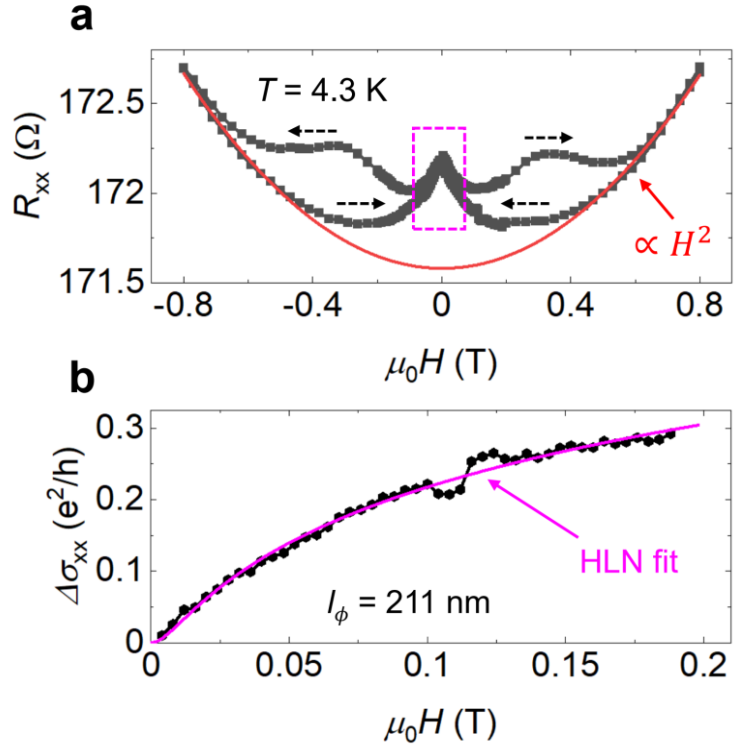

**Figure 6.** Longitudinal magnetoresistance and weak localization in a FGT/graphene heterostructure. (a) $R_{xx}$ versus $H$ measured at 4.3 K. The dashed arrows indicate the sweeping directions. The red line indicates a $H^2$-dependent positive magnetoresistance characteristic for carrier transport in graphene whereas the pink area highlights the resistance up-turn due to weak localization. (b) The magnetic-field induced variation in the longitudinal conductivity $\Delta\sigma_{xx}(H) = \sigma_{xx}(H) - \sigma_{xx}(0)$ can be well fitted by the HLN equation (pink line) in the low-field range during a down sweep of the external field. The obtained fitting parameters correspond to a phase coherence length of $l_\phi$ = 211 nm. The longitudinal conductivity has been calculated using the relation $\sigma_{xx} = 1/(R_{xx} \times W/L)$ with W and L being the sample width and contact length, respectively.

## SUMMARY

Our results demonstrate the successful growth of a large-area 2D heterostructure with high quality of the individual FGT and graphene components, as well as of the interface between them, which was previously achieved only by manually assembling flakes of the different materials. This constitutes a critical step towards the implementation of these materials in future technologies. Morphological and structural characterization revealed the fabrication of continuous heterostructure films with a sharp interface between FGT and graphene. Interestingly, stacking faults related to the presence of single vdW layers with thicknesses exceeding those expected for the $Fe_5GeTe_2$ phase are identified by STEM. We expect these to be FGT phases with Fe composition higher than 5 and potentially enhanced magnetic properties. Future investigations will be carried out to clarify this aspect. Different methods used to study the magnetic and transport properties revealed that the ferromagnetic order persists well above room temperature with perpendicular magnetic anisotropy. In addition, the underlying graphene film continues to show quantum features in its electronic properties, attesting to the conservation of its high quality after vdW epitaxy of FGT.

**ACKNOWLEDGEMENTS**

The authors would like to thank H.-P. Schönherr, C. Hermann, C. Stemmler, and K. Morgenroth for their dedicated maintenance of the MBE system. The authors also appreciate the critical reading of the manuscript by A. Hernández-Mínguez and S. Sadofiev. They also acknowledge the provision of beamtime under the project HC-4068 at the European Synchrotron Radiation Facility (ESRF), located in Grenoble (France), and at ALBA synchrotron in Spain under the project 2021035105. Dr. Juan Rubio-Zuazo and Dr. Jesus López-Sánchez are acknowledged for their assistance with the GID measurements. ICN2 acknowledges support from the European Union Horizon 2020 research and innovation programme under Grant Agreement No. 881603 (Graphene Flagship) and of the Spanish Research Agency (AEI), Ministry of Science and Innovation, MCIN/AEI/10.13039/501100011033, under grants PID2019-111773RB-I00 and CEX2021-001214-S (Severo Ochoa Centres of Excellence programme). A.I.F. acknowledges support of a fellowship from "la Caixa" Foundation (ID 100010434) with code LCF/BQ/DI18/11660030 and of H2020 Marie Skłodowska-Curie grant No. 713673. A.I.F. is a Serra Húnter Fellow.

**EXPERIMENTAL SECTION**

*Epitaxial graphene on SiC.* Semi-insulating 4H-SiC(0001) substrates with sizes of either 1.0 cm x 1.0 cm or 0.5 cm x 1.0 cm were chemically cleaned in n-butyl-acetate, acetone, and propanol. After that, they were loaded in an induction-heated furnace pumped down to a base pressure of $10^{-4}$ mbar. In order to remove surface irregularities and create well-defined stepped surfaces, an H-etching treatment was performed at 1400ºC for 15 min in forming gas atmosphere (95% Ar and 5% $H_2$) before graphene growth. Synthesis of epitaxial graphene was finally achieved via surface graphitization at a temperature of 1600ºC for 15 min in Ar atmosphere. Both H-etching and graphene growth were performed in a pressure of 900 mbar and a gas flow rate of 500 sccm. Further information on the formation of the epitaxial graphene on SiC and the influence of the SiC morphology can be found elsewhere.[30,53]

*MBE growth of $Fe_{5-x}GeTe_2$.* The 15 nm thick FGT films were grown in ultra-high vacuum (UHV, base pressure around 5 × $10^{-11}$ mbar) using high purity Fe, Ge, and Te evaporated from effusion cells at 1330 °C, 1050 °C, and 308 °C (hot-lip at 440 °C), respectively. The flux for each element was obtained by measuring the beam equivalent pressure employing a pressure gauge. Based on the film composition obtained by RBS analyses, the flux ratios were calibrated aiming at the growth of films with the nominal composition, i.e., $Fe_5GeTe_2$. The graphene/SiC(0001) templates were outgassed at 450 °C for at least one hour and then cooled to 300 °C for FGT growth. Prior to that, they were coated with 1 µm of Ti on the backside via electron beam evaporation to allow non-contact heating by radiation. *In-situ* growth monitoring was performed by reflection high energy electron diffraction (RHEED). Finally, the FGT films were capped *in-situ* with a ~ 5 nm thick Te layer deposited after sample cooling to room temperature. This procedure was adopted in order to avoid significant surface oxidation upon air exposure. Only the AFM, XRD, and RBS analyses were performed using uncapped samples. For AFM and XRD, the measurements could be started within a few minutes after the sample was unloaded from the MBE system. For the RBS analysis, the Te capping is unwanted in order to facilitate the compositional analyses. In this case, the samples were measured about three days after unloading from the MBE system.

*Rutherford backscattering spectrometry.* The composition of the samples was investigated by means of RBS using 1.4 MeV $He^+$ ions with a scattering angle of 170°. Their stoichiometry was determined by integration of the areal density from the well separated elemental signals of the different elements (Fe, Ge, Te).

*Atomic force microscopy.* The surface topography of the FGT films were investigated via standard tapping mode measurements performed in ambient conditions. The measurements were performed in uncapped samples within hours after MBE growth.

*X-ray diffraction.* Lab-based omega/2theta scans were recorded with a conventional X'Pert Pro MRD from Malvern Panalytical using Cu-Ka1 radiation ($\lambda$=1.54056 Å) as selected by a hybrid monochromator. The measurements were performed on uncapped samples within hours after MBE growth.

*Synchrotron-based grazing incident diffraction (GID).* This experiment was performed at the BM25-SpLine beamline at The European Synchrotron (ESRF) in Grenoble. An incidence angle of the illuminating X-rays of 0.2° sufficiently suppresses strong scattering by the substrate, and thus makes this method highly surface sensitive. An X-ray wavelength of 0.729 Å (corresponding to a photon energy of 17 keV), selected by a Si(111) monochromator, enables the inspection of a comparatively large area in reciprocal space, which is important for accessing multiple reflections of the same lattice plane family.

*Transmission electron microscopy.* The thin lamellas imaged in the TEM experiments were prepared utilizing a JEOL JIB-4501 focused ion-beam (FIB) microscope. A JEOL ARM 200F aberration corrected microscope was operated in scanning transmission electron microscopy (STEM) mode at 200 kV with a beam current of 68 pA. A high angle annular dark field detector (HAADF) was used to acquire the images with inner and outer collection angles of 54 and 220 mrad, respectively, whereas the bright field detector had a collection angle of 36 mrad.

*Magnetotransport characterization.* The magnetotransport measurements were carried out on a rectangular strip of the FGT/graphene heterostructure bonded on a chip carrier with Al contact wires. A constant current of $I_{xx}$ = 200 mA was applied along the long side of the strip (x-direction) through two contacts close to the edges of the strip. For the determination of the longitudinal resistance, the voltage ($V_{xx}$) between two inner contacts was measured. The Hall voltage ($V_{xy}$) was measured in the orthogonal direction (y-direction) between two contacts close to the center of the strip. The measurements were performed at temperatures between 4.3 and 400 K under vacuum conditions ($10^{-6}$ to $10^{-5}$ mbar) with an external perpendicular (z-direction) magnetic field of up to 0.8 T.

*Superconducting quantum interference device (SQUID) magnetometry.* The magnetometry measurements were performed by a Quantum Design MPMS 5XL SQUID at various temperatures between 2 and 375 K with an applied magnetic field of up to 4.5 T along both out-of-plane (OP) and in-plane (IP) directions, using the same sample.

*X-Ray absorption spectroscopy (XAS) and X-ray magnetic circular dichroism (XMCD).* XAS spectra at the Fe $L_3$ and $L_2$ edges were recorded in BL-29 (BOREAS) beamline at the ALBA synchrotron (Spain), which provides an ultra-high-vacuum sample environment with a base temperature of ∼ 3 K and a magnetic field $B$ of up to 6 T. Measurements used total-electron-yield detection, where the drain current was measured from the sample to the ground. The magnetic field was applied along the X-ray beam both at normal and grazing incidence (20º) relative to the sample plane in order to obtain information on the out-of-plane and in-plane magnetization, respectively. XMCD was obtained by calculating the difference between XAS spectra obtained with the photon helicity vector antiparallel and parallel to the magnetic field. (see Supporting Information).

# SUPPORTING INFORMATION

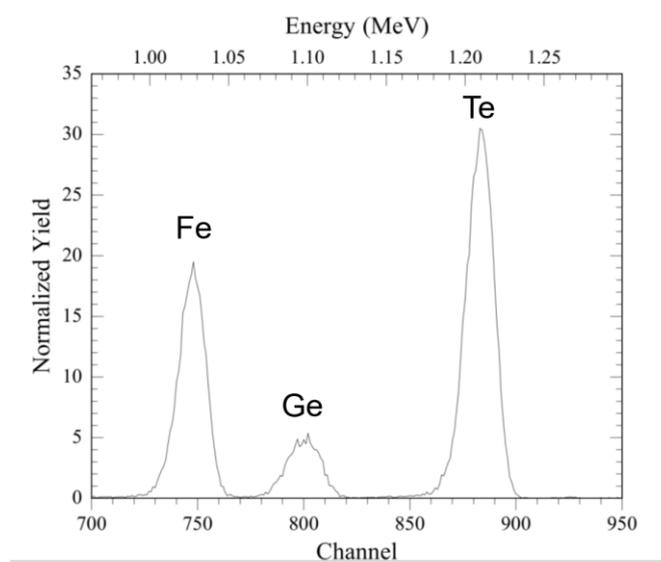

Figure S1. Exemplary Rutherford backscattering spectrometry (RBS) curve for an uncapped, 15 nm FGT film. The stoichiometry is determined via the RUMP simulation, which considers the integration of the areal density from the well-separated elemental signals of the different elements (Fe, Ge, Te).

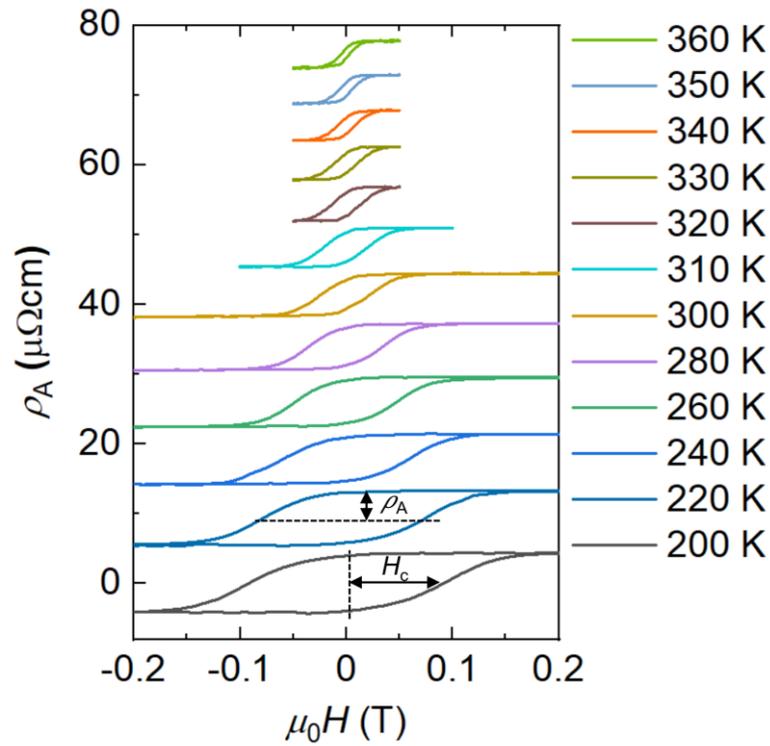

Figure S2. AHE curves after subtraction of the linear OHE contribution. The definition of the quantities $\rho_A$ and $H_C$ are indicated by double arrows.

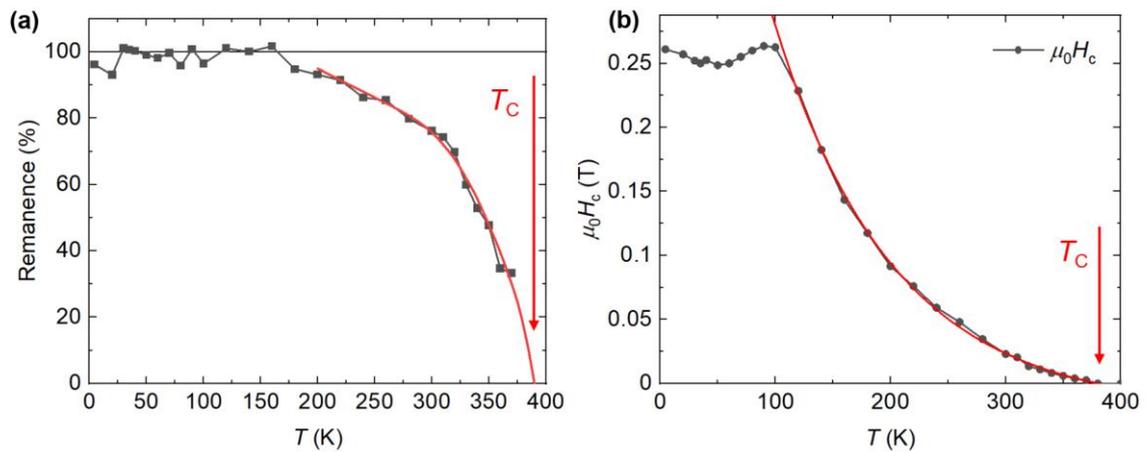

Figure S3. Curie temperature $T_C$ estimated from the temperature decay of (a) the AHE remanence and (b) the coercivity $H_C$. The red lines are shown as guides to the eye. The obtained values are consistent with the results extracted by the Arrott-plot analysis (see Figure 3 in the main text).

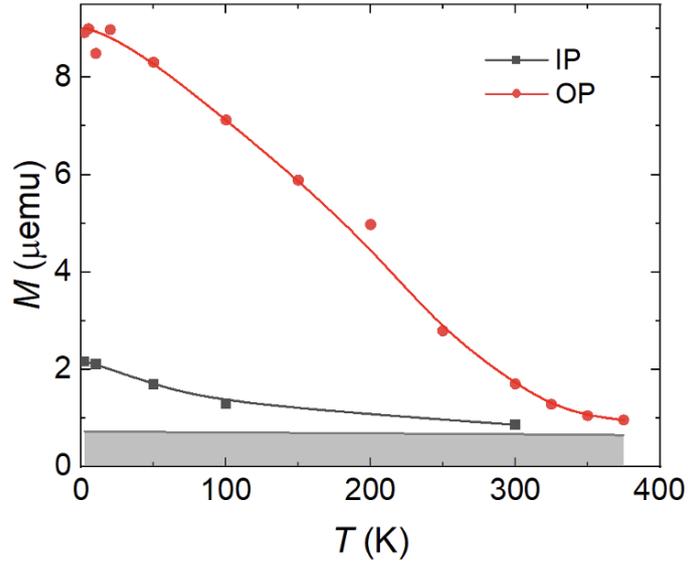

Figure S4. Temperature dependence of remanent magnetization obtained by SQUID magnetometry. The gray area indicates the background signal originating from the SiC substrate determined from paramagnetic magnetization curves obtained from a pure substrate piece measured at both room temperature and at 10 K.

XAS/XMCD Analyses

A sum-rules analysis was performed in the way developed by Thole *et al.*[1] that allows us to extract the atomic orbital and spin magnetic moment from the XAS and XMCD spectra. The different absorption and dichroic integrals and sum rules relationships are defined as follows:

$$r = \int_{L_3+L_2} \left(\frac{\sigma^+ + \sigma^-}{2} - \sigma_b\right) dE \quad \langle L_z \rangle = -\frac{N_h}{P\cos(\theta)} \frac{2q}{3r}$$

$$p = \int_{L_3} (\sigma^+ - \sigma^-) dE \quad \langle S_{eff} \rangle = \langle S_z \rangle + \frac{7}{2}\langle T_z \rangle = -\frac{N_h}{P\cos(\theta)} \frac{3p-2q}{2r}$$

$$q = \int_{L_3+L_2} (\sigma^+ - \sigma^-) dE \quad \langle M_{eff} \rangle = \langle L_z \rangle + 2\langle S_{eff} \rangle$$

where $\sigma^+$ ($\sigma^-$) is the absorption spectra recorded with right (left) circularly polarized X-rays, $\sigma_b$ is the double step function background of secondary electrons, $L_z$ and $S_z$ are the atomic orbital and spin magnetic moment, $S_{eff}$ is an effective spin moment extracted from the sum-rules analysis that includes the intra-atomic spin dipole operator $T_z$, $N_h$ is the number of holes in the $3d$ band, $P$ is the degree of circular polarization of the X-rays (close to 1, i.e. 100%, at photon energies around the Fe L3,2 edges in BOREAS beamline) and $\theta$ is the angle between the magnetization and X-ray beam, assumed to be zero under the maximum applied magnetic field with a strength of 6 T.

We took the number of holes equal to $N_h = 3.65$, as calculated for FGT,[2] and summarized the results of the sum rules analysis at T = 3 K in the table shown in Figure S5. The difference between the effective spin moments calculated in normal and grazing incidence is likely due to the anisotropy of the intra-atomic dipole operator.[3]

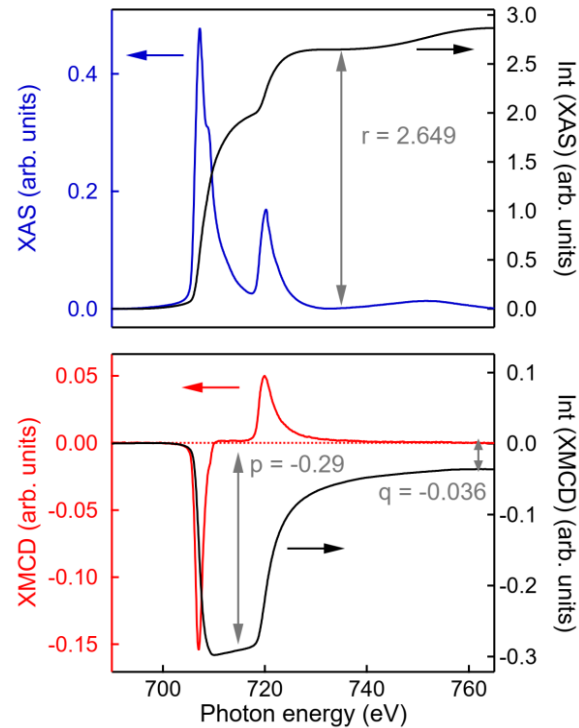

|  | Lz ($\mu_B$) | Seff ($\mu_B$) | Meff ($\mu_B$) |
|---|---|---|---|
| Normal inc. | 0.03 | 0.55 | 1.13 |
| Grazing inc. | 0.03 | 0.5 | 1.03 |

Figure S5. Normalized XAS and XMCD measurements. (top) Average XAS corrected by the double step background, together with its integral (r). (bottom) XMCD calculated as the difference between spectra measured using left-circularly and right-circularly polarized x-rays. The integral of XMCD is also included. Arrows mark the photon energy at which the integral of the $L_3$ (p) and the $L_2+L_3$ (q) are taken for the sum rules analysis. The measurements were done at a sample temperature of 3 K and under an external field of 6 T applied along the direction perpendicular to the sample surface, collinear with the beam direction. The table summarizes the results of the sum rules analysis at T = 3 K.

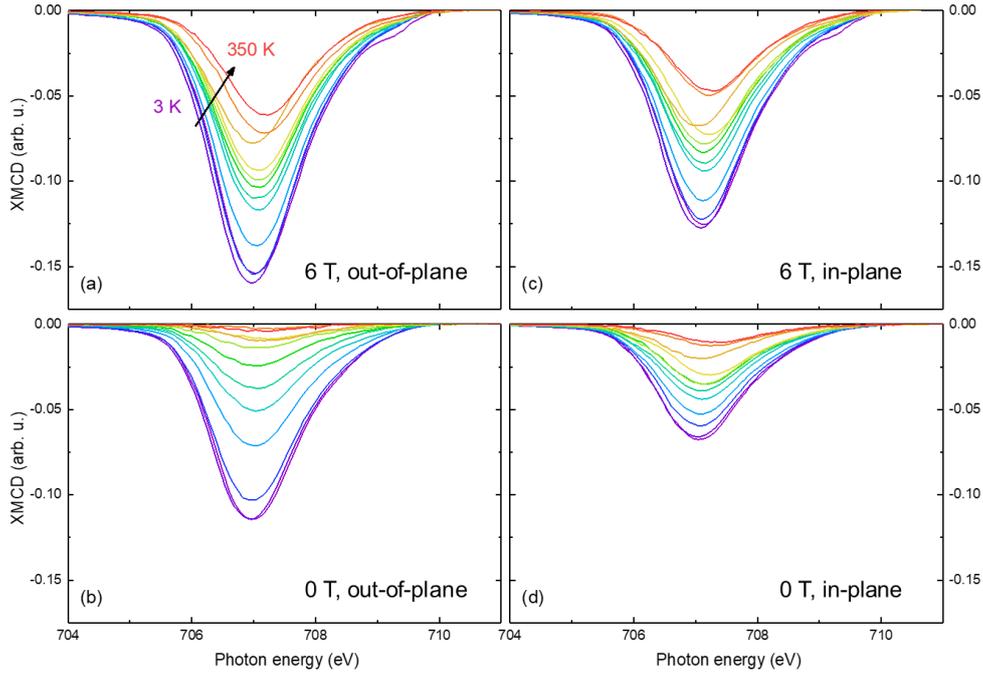

Figure S6. Closeup of the Fe $L_3$ XMCD spectra recorded at different temperatures between 3 and 350 K. The measurements were performed under an external magnetic field of (a) 6 T and (b) 0 T, applied along the direction perpendicular to the sample surface, and of (c) 6 T and (d) 0 T, applied at 20° from the sample plane. In both configurations, the magnetic field is applied collinearly with the beam direction.

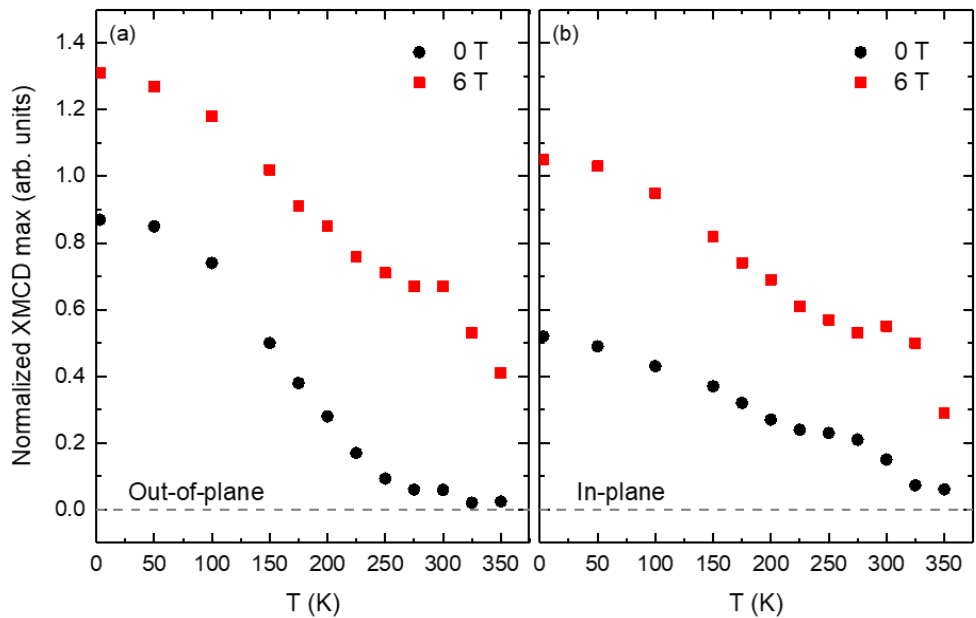

Figure S7. Temperature dependence of the Fe $L_3$ XMCD maximum in remanence (0 T, black circles), and saturation (6 T, red squares). The external magnetic field is applied along the direction perpendicular to the sample surface (out-of-plane, left panel) and at 20° from the sample plane (in-plane, right panel). XMCD signal was normalized to the Fe $L_3$ XAS maximum.